\documentclass[twocolumn,amsmath,nosuperscriptaddress,amssymb,nopacs,nofootinbib,prx]{revtex4-1}

\usepackage[usenames,dvips]{color}
\usepackage{graphicx}
\usepackage{dcolumn}
\usepackage{bm}
\usepackage{mathtools}
\usepackage{epstopdf}
\usepackage{esint}
\usepackage{xcolor}
\usepackage{dsfont}
\usepackage{nicefrac}
\usepackage[normalem]{ulem}

\usepackage[colorlinks=true,citecolor=blue,linkcolor=magenta, urlcolor=magenta]{hyperref}

\newcommand{\bea}{\begin{eqnarray}}
\newcommand{\eea}{\end{eqnarray}}
\newcommand{\bt}{\textbf}

\newcommand{\phd}{\phantom{\dag}}
\newcommand{\ph}{\phantom{.}}

\newcommand{\noi}{\noindent}
\newcommand{\no}{\nonumber}

\begin{document}
\def\v#1{{\bf #1}}

\title{Diagnosing Topological Phase Transitions in 1D Superconductors\\ using Berry Singularity Markers}

\author{Panagiotis Kotetes}
\email{kotetes@itp.ac.cn}
\affiliation{CAS Key Laboratory of Theoretical Physics, Institute of Theoretical Physics, Chinese Academy of Sciences, Beijing 100190, China}

\vskip 1.cm
\begin{abstract}
In this work I demonstrate how to characterize topological phase transitions in BDI symmetry class superconductors (SCs) in 1D, using the recently introduced approach of Berry singularity markers (BSMs). In particular, I apply the BSM method to the celebrated Kitaev chain model, as well as to a variant of it, which contains both nearest and next nearest neighbor equal spin pai\-rings. Depending on the situation, I identify pairs of external fields which can detect the topological charges of the Berry singularities which are responsible for the various topological phase transitions. These pairs of fields consist of either a flux knob which controls the supercurrent flow through the SC, or, strain, combined with a field which can tune the chemical potential of the system. Employing the present BSM approach appears to be within experimental reach for topological nanowire hybrids.
\end{abstract}

\maketitle

\section{Introduction}

A little more than a decade after the beginning of the intense pursuit of Majorana zero modes (MZMs) in nanowire hybrids~\cite{AliceaPRB,LutchynPRL,OregPRL}, a tremendous progress has been recorded in various directions, e.g., the theoretical concepts, the material fabrication, the measurement protocols and, last but not least, the claims for the discovery of MZM fin\-ger\-prints in a number of expe\-ri\-ments~\cite{Mourik,MTEarly,Das,MT,Sven,HaoZhang}. For an overview, see for instance Refs.~\onlinecite{LutchynNatRevMat,PradaRev}. Notably, in most cases, the detection of MZMs has been pursued by identifying a zero-bias peak (ZBP) in the differential tunneling conductance. Specifically, the observation of a ZBP with a quantized height equal to $2e^2/h$~\cite{LawZBP,FlensbergZBP}, has been linked to the presence of a MZM. However, such a signature is not unique to MZMs~\cite{PradaRev}. Indeed, the retraction~\cite{HaoZhang2} of the initial claim for a quantized ZBP~\cite{HaoZhang} in InSb hybrids, along with a number of recent experimental observations~\cite{FrolovUbiquitous,YuQuasiMZM,Katsaros}, reveal that near-zero-energy topologically-trivial Andreev bound states are capable of reconciliating the arising phe\-no\-me\-no\-lo\-gy without invo\-king MZMs~\cite{Brouwer,Akhmerov,JLiu,Bagrets,Prada2012,DimaPeak,RoyTewari,SauSarma,PinningInteractions}. All the above have recently spurred a strong scepticism re\-gar\-ding the discovery of MZMs in nanowire hybrids and the perspectives of this search.

Motivated by the above, a number of theoretical works have appeared recently~\cite{CXLiu,HellInterferometry,Moore,CXLiu2,Reeg,Woods,Vuik,PanDasSarma}, which inve\-sti\-ga\-te the role of disorder in such systems, and discuss potential approaches that may quantify the topological quality and visibility that characterizes a hybrid device. Even more, several groups have proposed protocols for evaluating the visibility of the system and the topological character of the bulk energy gap closing by investigating nonlocal responses~{\color{black}\cite{Roshdal,Jeroen,Menard3Terminal,3Terminal,PikulinProtocol,Hess}}. Nonetheless, up to my know\-ledge, none of the above approaches has suggested to infer the topological properties of the device by identifying the topological charges of the Berry singularities which are responsible for the various topological phase transitions. 

In this Article, I propose to quantify the visibility of the target {\color{black}topological superconductor (TSC)}, and diagnose the topological character of the arising phase transitions by em\-ploying Berry singularity markers (BSMs). The BSMs, which were recently introduced in a personal work~\cite{PK}, are constructed using (quasi)equilibrium responses of the target system, which are measured in the presence of suitable external fields. The latter need to exhibit a slow spatial/temporal va\-ria\-tion compared to the cha\-racte\-ri\-stic length/time scales of the system. By construction, BSMs becomes sizeable and tend to topological invariant quantities only when Berry singularities are detected. Moreover, they yield the charge of the arrested singularity and, therefore, enable the topological characterization of a gap clo\-sing point and/or a phase transition. More importantly, inferring the BSM can be implemented in a lock-in fashion~\cite{PK}, which is a procedure that can render the eva\-lua\-tion of the BSM immune to the influence of noise and disorder.

\begin{figure}[t!]
\begin{center}
\includegraphics[width=0.85\columnwidth]{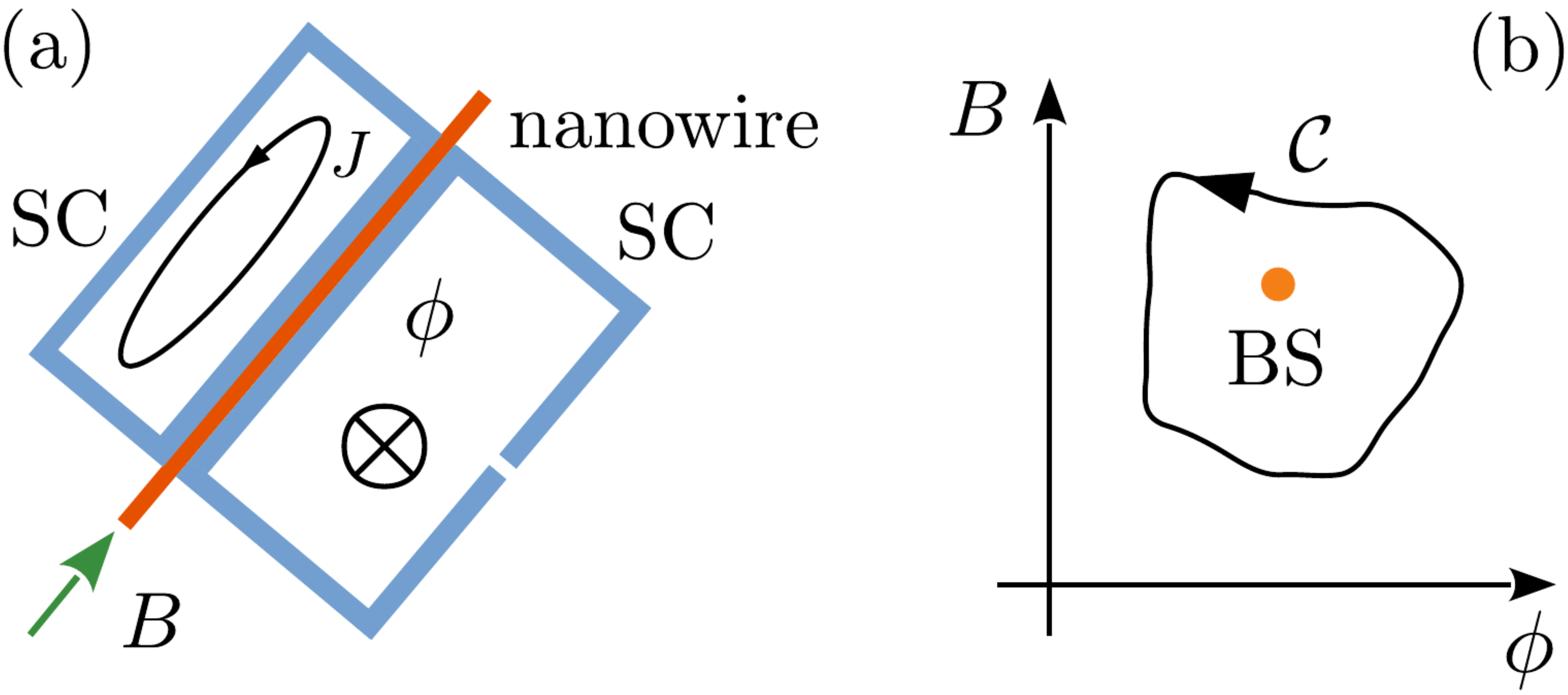}
\end{center}
\caption{(a) Sketch of a topological hybrid based on a Rashba semiconductor nanowire in proximity to two loops made of superconductors (SCs). The nanowire feels a Zeeman field $B$. One loop is open and pierced by flux which yields a phase gra\-dient $\phi$ per unit length of the proximitized nanowire. A supercurrent $J$ flows through the nanowire which can be detected through the flux induced in the closed loop. (b) The Berry singularity (BS) marker is obtained by spatiotemporally varying $(\phi,B)$ along a closed loop ${\cal C}$ encircling a gap closing point of the bulk energy spectrum and its related BS.} 
\label{fig:Figure1}
\end{figure}

In the present manuscript, I first extend the framework that was developed in Ref.~\onlinecite{PK} for AIII to\-po\-lo\-gi\-cal insulators, to the case of 1D $p$-wave {\color{black}superconductors (SCs)} which belong to the BDI symmetry class. {\color{black}Notably, despite the fact that AIII and BDI classes share a number of common features, they are yet sufficiently distinct, so to render the extensions discussed here more than a mere application of my previous work. In more detail, the analysis of the present manuscript} focuses on the Kitaev chain model~\cite{KitaevUnpaired}, a continuum version of which, is predicted to effectively describe~\cite{AliceaTQC} the low-energy physics of the semiconductor nanowire hybrids~\cite{LutchynPRL,OregPRL}. As I show, the already applied magnetic field in the nanowire and a flux which induces a supercurrent flow through it, map the Berry sin\-gu\-la\-ri\-ties to a synthetic space, and serve as the two external fields which are here required for the evaluation of the BSM. See Fig.~\ref{fig:Figure1} for a schematic.

Apart from demonstrating the implementation of the BSM theory in the case of the pristine Kitaev chain model, I consider further extensions which can be re\-le\-vant for experiments under the assumption of a BDI class~\cite{Ryu}. Spe\-ci\-fi\-cal\-ly, by here allowing for long-ranged pai\-ring terms, I address the case of additional Berry sin\-gu\-la\-ri\-ties which come in pairs and possesss identical topological charges. Such scenarios become relevant for effective nanowire models~\cite{Brydon,Li,HeimesInter} which describe coupled Yu-Shiba-Rusinov states in topological ferromagnetic chains~\cite{Yazdani1,Yazdani2,Pawlak}. The presence of the additional Berry singularities can either reduce the visibility of the BSM or narrow down the window of the bulk insulating regime. Notably, as I show, effecting strain and controlling the magnetization field, allows to construct alternative BSMs that can be used to detect the Berry singularity pairs. 

The remainder is organized as follows. Section~\ref{sec:SectionII} introduces the general model Hamiltionian to be discussed here. Sections~\ref{sec:SectionIII} and~\ref{sec:SectionIV} contain the main analysis of the topological, response and BSM properties of the standard and extended Kitaev chain models, respectively. Finally, Sec.~\ref{sec:SectionV} discusses conclusions and open questions.  

\section{Model Hamiltonian}\label{sec:SectionII}

The various topological scenarios discussed in this work are described in terms of the following Hamiltonian operator, which is defined using the framework of second quantization, and reads as:
\begin{align}
\hat{H}=\big(B-u\big)\sum_nc_n^\dag c_n+\frac{u}{2}\sum_n\Big(c_n^\dag c_{n+1}+{\rm h.c.}\Big)\no\\+\frac{1}{4}\sum_{n,m}\Big(c_n^\dag\Delta_{nm}c_m^\dag-c_n\Delta_{mn}^*c_m\Big)\,.
\label{eq:ModelHamiltonian}
\end{align}

\noi $c_n$/$c_n^\dag$ defines the annihilation/crea\-tion ope\-ra\-tor of a spinless electron at lattice site $n$. The lattice constant is set to unity throughout. $B$ controls the chemical potential while the pa\-ra\-me\-ter $u$ defines the strength for nearest neighbor hopping. The nonlocal term $\Delta_{nm}$ denotes the coordinate space superconducting gap, which pairs up electrons residing on the different lattice sites $n$ and $m$. In the remainder, all energy scales are expressed in terms of $u$, which I set to unity from now on. 

To proceed, I consider the following concrete structure for the pairing term:
\begin{align}
\Delta_{nm}=e^{-i\varphi_n/2}\Delta_{n-m}e^{-i\varphi_m/2}
\end{align}

\noi with $\varphi_n$ denoting the global or ``center-of-mass'' superconducting phase. Besides this overall local phase, the re\-mai\-ning part $\Delta_{n-m}$ of the pairing term $\Delta_{nm}$ is here assumed to depend only on the differences between the various lattice sites involved. As a result, it is odd under inversion, i.e., $\Delta_{m-n}=-\Delta_{n-m}$. To proceed, it is convenient to gauge away the phase factors $e^{-i\varphi_n/2}$ by means of the gauge transformation $c_n\mapsto e^{-i\varphi_n/2}c_n$. This manipulation transfers the phase factor of the pairing gap to the hopping term, which now becomes:
\begin{align}
c_n^\dag c_{n+1}\mapsto {\rm Exp}\left[-i\big(\varphi_{n+1}-\varphi_n\big)/2\right]c_n^\dag c_{n+1}\,.
\end{align}

The spatially varying superconducting phase is induced by the bond-defined flux element $\phi_{n+1;n}\in[-\pi,\pi)$: $\phi_{n+1;n}=(\varphi_{n+1}-\varphi_n)/2$. This flux pierces an almost-closed superconducting loop that terminates at the sites $n$ and $n+1$ of the SC, and induces a spatial gradient in the superconducting phase. Note that the loop is embedded in the 2D coordinate space, as depicted in Fig.~\ref{fig:Figure1}. In the present work, I restrict to cases in which the flux is either constant or varies sufficiently slow in space/time, so that a spatiotemporal adia\-ba\-tic approach applies. Given this assumption, both si\-tua\-tions can be described by a spatially (quasi)constant $\phi$, which effectively renders the model Hamiltonian invariant with respect to both space and time translations. 

Within the adiabatic picture and the emergent translational invariance assumed, it is convenient to perform a lattice Fourier transform $c_n=\nicefrac{1}{\sqrt{N}}\sum_{k}e^{ik n}c_k$ and describe the bulk properties of the system in the first Brillouin zone defined by the wavenumber $k\in[0,2\pi)$. $N$ defines the number of sites which comprise the chain. After introducing the $c_k$ and $c_k^\dag$ operators, I addi\-tio\-nal\-ly perform a unitary transformation to the Majorana basis~\cite{KitaevUnpaired} to establish a {\color{black}connection} to the Hamiltonian employed in~\cite{PK}. The emerging Majorana ope\-ra\-tors $\gamma_k$ and $\tilde{\gamma}_k$ are defined through $c_k=(\gamma_k+i\tilde{\gamma}_k)/\sqrt{2}$. These describe charge neutral objects, which is a property reflected in the relations $\gamma_k^\dag=\gamma_{-k}$ and $\tilde{\gamma}_k^\dag=\tilde{\gamma}_{-k}$. In addition, they fulfill the anticommutators $\{\gamma_q,\tilde{\gamma}_k\}=0$ and $\{\gamma_q,\gamma_{-k}\}=\{\tilde{\gamma}_q,\tilde{\gamma}_{-k}\}=\delta_{q,k}$ for all $k,q\in[0,2\pi)$.

With the use of the above formalism, the model of Eq.~\eqref{eq:ModelHamiltonian} is compactly written as:
\begin{align}
\hat{H}(\phi,B)=\frac{1}{2}\sum_k\bm{\Gamma}_{-k}^\intercal\hat{H}_{\rm BdG}(k,\phi,B)\bm{\Gamma}_k
\end{align}

\noi with the multicomponent Majorana operator being defined as $\bm{\Gamma}_k^\intercal=\big(\gamma_k\,,\tilde{\gamma}_k\big)$ where $^\intercal$ effects matrix transposition. The matrix $\hat{H}_{\rm BdG}(k,\phi,B)$ defines the Bogoliubov - de Gennes (BdG) Hamiltonian. For the upcoming ana\-ly\-sis I consider the $k$-space pairing term:
\begin{align}
\Delta(k)=i\sin k\big(1-\beta\cos k\big),
\end{align}

\noi which is defined by means of the lattice Fourier transform $\Delta(k)=\sum_ne^{-ik(n-m)}\Delta_{n-m}$. In the above, I consider contributions from the two har\-mo\-nics $\sin k$ and $\sin(2k)$ which lead to nearest and next nearest neighbor Cooper pairing. {\color{black}While the} coefficient of each harmonic is ge\-ne\-ral\-ly complex, here the two pairing coefficients are chosen to see a zero relative phase difference. In the hybrid devices of interest, the spinless triplet pairing is induced by the combination of Rashba-type spin-orbit coupling (SOC) and spin-singlet superconductivity, in the additional presence of a magnetic field which polarizes the electron spin. Thus, the two harmonics can be viewed as the result of two SOC terms of a dif\-fe\-rent range. Such a si\-tua\-tion occurs for to\-po\-lo\-gi\-cal magnetic chains in ge\-ne\-ral, where the effective low-energy Hamiltonian contains both long-ranged hoppings and pairing terms~\cite{Nakosai, Pientka}. 

Putting all the above ingredients together, I express the BdG Hamiltonian compactly as:
\begin{align}
\hat{H}_{\rm BdG}(k,\phi,B)=-\sin\phi\sin k\mathds{1}_\tau+\bm{g}(k,\phi,B)\cdot\bm{\tau}\,. 
\label{eq:BdGHamiltonian}
\end{align}

\noi $\mathds{1}_\tau$ and $\bm{\tau}$ correspond to the unit and Pauli matrices defined in the Nambu $\tau$ space spanned by $\gamma_k$ and $\tilde{\gamma}_k$. The definition of the BdG Hamiltonian becomes complete after specifying the $\bm{g}(k,\phi,B)$ vector which takes the form: 
\begin{align}
\bm{g}=\big(\sin k(1-\beta\cos k),1-\cos\phi\cos k-B,0\big).\label{eq:gVector}
\end{align}

\noi The system is characterized by two energy bands, with their dispersions given by:
\begin{align}
E_\pm(k,\phi,B)=-\sin\phi\sin k\mathds{1}_\tau\pm|\bm{g}(k,\phi,B)|\,.
\end{align}

\noi One observes that when the flux is tuned to either one of its two time-reversal invariant (TRI) values $\phi=\{0,\pi\}$, the two dispersions satisfy $E_+(k,\phi,B)=-E_-(k,\phi,B)$ and, thus, the mo\-di\-fi\-ca\-tion of the various parameters can only lead to band touchings. Instead, when con\-si\-de\-ring fluxes away from the two TRI values, the two dispersions satisfy $E_+(k,M,\phi,B)=-E_-(-k,M,\phi,B)$, which is a property that unlocks the possibility of band crossings.

\section{Standard p-wave SC Model}\label{sec:SectionIII}

My analysis starts with the exploration of the properties of the standard model for a spinless $p$-wave SC~\cite{KitaevUnpaired}, which is obtained from Eq.~\eqref{eq:BdGHamiltonian} by setting $\beta=0$.

\subsection{Topological phases and Berry singularities}\label{Sec:TopoPhasesStandard}

When $\phi=\{0,\pi\}$, the Hamiltonian lies in class BDI for arbitrary va\-lues chosen for $B$~\cite{Ryu}. The topological properties of a BDI class SC are inferred by evaluating the winding number $w\in\mathbb{Z}$, which here simply reads as:
\bea
w(\phi,B)=\int_{\rm BZ}\frac{dk}{2\pi}\sum_{n,m}\varepsilon_{nm}\hat{g}_n(k,\phi,B)\partial_k\hat{g}_m(k,\phi,B)\,,\qquad\label{eq:WindingNumber}
\eea

\noi where $n,m=1,2$, $\varepsilon_{12}=-\varepsilon_{21}=1$, and $\hat{\bm{g}}=\bm{g}/|\bm{g}|$. The above topological invariant predicts that in the case of an open chain, a number of $|w|$ MZMs appear at each edge. These remain uncoupled due to the protection provided by the chiral symmetry effected by the operator $\Pi=\tau_3$.

Class D Hamiltonians set in for $\phi\neq\{0,\pi\}$. In principle, the change of the symmetry class also implies a change in the MZM phenomenology. Specifically, only a single MZM per edge is now accessible, and its to\-po\-lo\-gi\-cal protection is determined by the so-called Majorana number which constitutes a $\mathbb{Z}_2$ inva\-riant~\cite{KitaevUnpaired} ta\-king two possible values, say, $\pm1$. This inva\-riant changes sign when an odd number of band touchings take place. Notably, the Majorana number is meaningful only in the presence of a full bulk energy gap~\cite{KitaevUnpaired}. Hence, further attention needs to be paid in class D scenarios, since in this case band crossings become possible, which lead to gapless phases and robust Bogoliubov-Fermi points~\cite{ChiuRMP}.

It is important to remark that inducing a symmetry class transition BDI$\rightarrow$D via a non-TRI flux gives rise to a very special realization of a class D Hamiltonian. As one infers from Eqs.~\eqref{eq:BdGHamiltonian} and~\eqref{eq:gVector}, considering $\phi\neq\{0,\pi\}$ has a twofold effect. First, it mo\-di\-fies the strength of the hopping term and, second, it adds a diagonal term which generates net momentum. However, a dia\-go\-nal term cannot modify the eigenvectors of the BdG Hamiltonian and, in turn, it cannot qualitatively alter the topological behaviour of the sy\-stem. This is of course true as long as the system does not undergo a band tou\-ching or cros\-sing. Indeed, such non-TRI fluxes are known to lead to gapless phases~\cite{Romito}. Hence, by restricting to the insu\-lating regime, a flux $\phi\neq\{0,\pi\}$ does not inva\-li\-da\-te the win\-ding number which can be still used to classify the topological phases of the system. Consequently, a sufficiently weak flux appears to provide a non-invasive knob to probe the topological properties of a BDI class Hamiltonian.

For the given choice of parameter values in Eq.~\eqref{eq:BdGHamiltonian}, gapless phases do not appear, unless $\beta$ is nonzero. Therefore, the topological properties of the standard model di\-scussed in this section can be fully described by the win\-ding number defined in Eq.~\eqref{eq:WindingNumber}. One ends up with two topologically-nontrivial phases with $w=\pm1$. The associated phase transitions occur along two paths in $(\phi,B)$ space, which stem from the gap closings at $k=0$ and $k=\pi$, and are given by the expressions: $B=1-\cos\phi$ and $B=1+\cos\phi$, respectively. 

The total topological charge of the Berry singularities responsible for a topological phase transition is obtained by the difference of the winding number across the respective transition. Here, every topological phase transition involves only a single Berry singularity. The topological charge $Q_s$ of the $s$-th singularity is given by the vorticity:
\begin{align}
Q_s(\phi)=-\ointctrclockwise_{{\cal C}_{\bm{K}_s}}\frac{d\bm{K}}{2\pi}\cdot\sum_{n,m}\varepsilon_{nm}\hat{g}_n(\bm{K},\phi)\partial_{\bm{K}}\hat{g}_m(\bm{\bm{K}},\phi)\,,
\label{eq:Vorticity}
\end{align}

\noi where $n,m=1,2$. In the above I introduced the synthetic space $\bm{K}=(k,B)$. ${\cal C}_{\bm{K}_s}$ defines a closed loop which encircles the singularity located at $\bm{K}_s$. Straightforward calculations for $\phi=0$, yield that the vorticities of the Berry singularities at $k=\{0,\pi\}$ equal to $Q=\{1,-1\}$. 

\subsection{Response to the external fields}

The construction of the BSM discussed in Ref.~\onlinecite{PK} requires experimentally inferring and theoretically pre\-dic\-ting the conjugate fields $(J,M)$ of $(\phi,B)$, which are defined through the relations:
\begin{align}
J(\phi,B)=-\partial_\phi E(\phi,B)\quad{\rm and}\quad M(\phi,B)=-\partial_B E(\phi,B),
\end{align}

\noi where $E(\phi,B)$ defines the total energy of the system at zero temperature ($T$). In general, $E$ receives contributions from both $E_\pm(k,\phi,B)$ bands, even in the $T=0$ limit, due to the possible band crossings which may arise for $\phi\neq\{0,\pi\}$. The conjugate fields $J$ and $M$ correspond to a supercurrent flowing through the TSC and the particle density, respectively. Note that in engineered TSCs, $B$ is associated with a magnetic field. Thus, in these cases, $M$ corresponds to the magnetization of the sy\-stem. The conjugate fields possess similar symmetry properties to the components of the $\bm{g}$ vector entering the Hamiltonian and this similarity allows the construction of the BSM.

It is instructive to alternative view the response fields as the expectation values of corresponding microscopic operators defined in terms of the Hamiltonian. In the present case, $J$ and $M$ originate from the operators:
\bea
\hat{J}(k,\phi,B)&=&-\partial_\phi\hat{H}(k,\phi,B)\no\\
&=&\cos\phi\sin k\hat{\mathds{1}}_\tau-\sin\phi\cos k\tau_2\,,\label{eq:Jdef}\\
\hat{M}(k,\phi,B)&=&-\partial_B\hat{H}(k,\phi,B)=\tau_2\,.
\eea

\noi When there is a full gap in the spectrum only $E_-(k,\phi,B)$ contributes to the expectation values of the above ope\-ra\-tors. Even more, the presence of the built-in charge-conjugation symmetry $\Xi$ of the BdG Hamiltonian, which in the given Majorana basis is effected by the complex conjugation ${\cal K}$, implies that the contribution of the dia\-go\-nal $\cos\phi\sin k\hat{\mathds{1}}_\tau$ term to the current is identically zero. Therefore, in this case, $J$ solely stems from the contribution of the $g_2(k,\phi,B)$ component. These two different terms give rise to the so-called paramagnetic and diamagnetic contributions to the supercurrent. The former (latter) is independent of $\phi$ (proportional to $\phi$) for $\phi\approx0$.

\begin{figure}[t!]
\begin{center}
\includegraphics[width=\columnwidth]{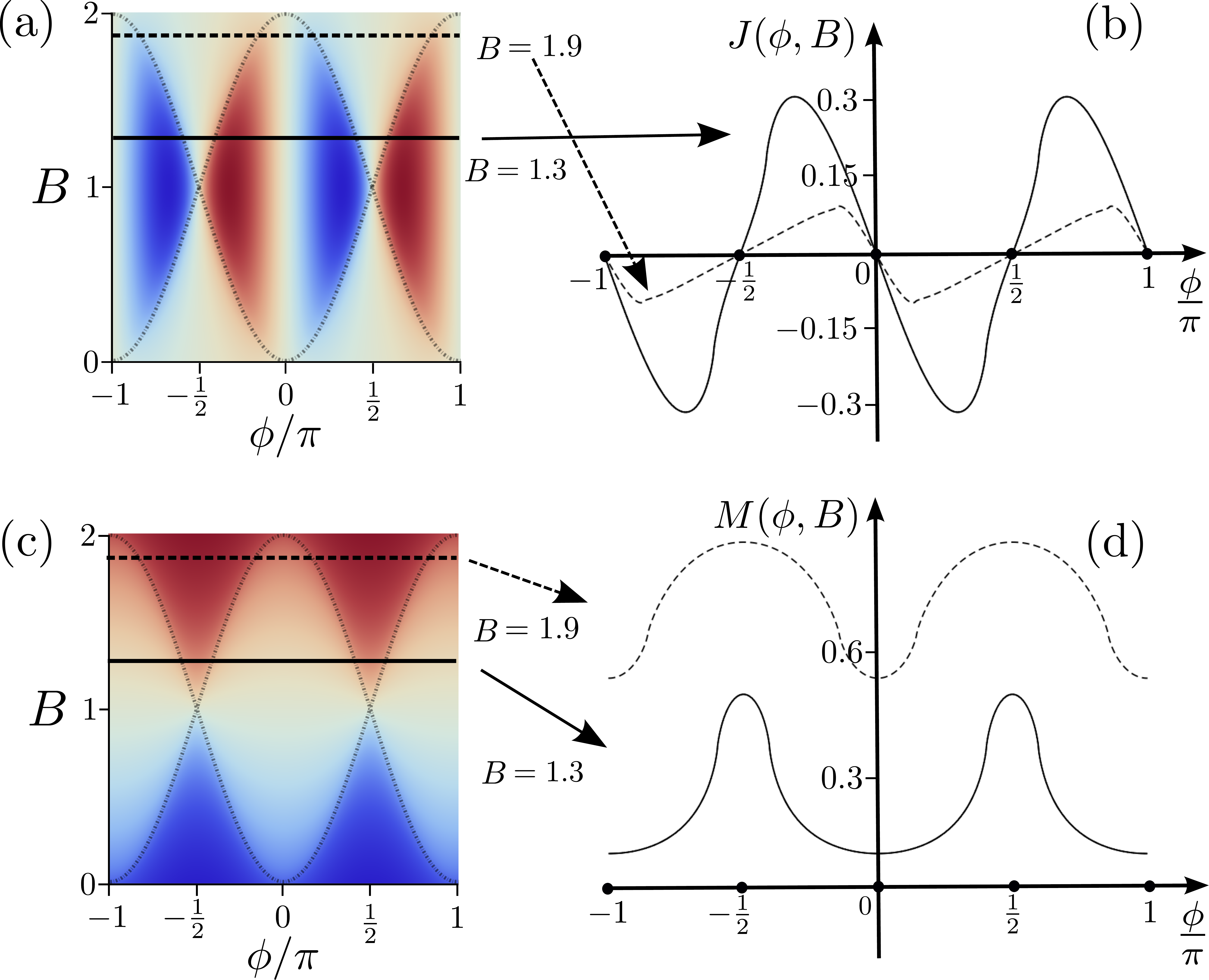}
\end{center}
\caption{$(J,M)$ as functions of their conjugate fields $(\phi,B)$, for the model of Eq.~\eqref{eq:BdGHamiltonian}, for $\beta=0$. Panels (a) and (c) depict these dependences using heat maps with a colorcoding which varies continuously between deep red to deep blue (\includegraphics[scale=0.06]{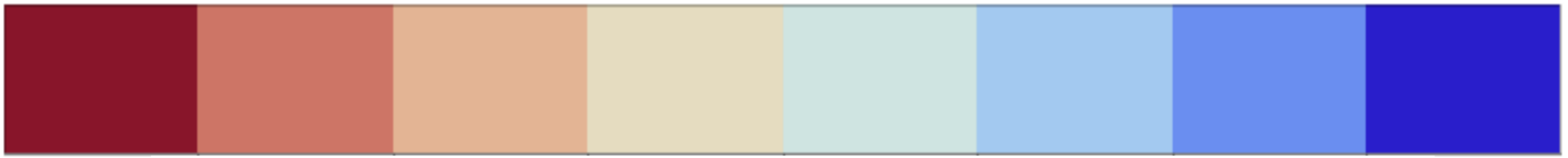}), with these two values corresponding one-to-one to the maximum and minimum of the depicted function. In both panels, I have superimposed the contour plots $B=1\pm\cos\phi$ (dot-dashed lines), which yield the locations of the Berry singularities in the synthetic $(\phi,B)$ space. Panels (b) and (d) are corresponding cuts of (a) and (c) for $B=1.9$ and $B=1.3$.}
\label{fig:Figure2}
\end{figure}

Figure~\ref{fig:Figure2} provides information regarding the behaviour of $(J,M)$ for varying $(\phi,B)$. From this figure, one first indeed confirms that the conjugate fields possess the desired symmetry properties, so that a correspondence between $(g_1,g_2)\rightarrow(J,M)$ and $k\rightarrow\phi$ is established. In more detail, panels \{(a),\,(c)\} depict heat maps of the response fields, while panels \{(b),\,(d)\} show cuts of the former for the two values $B=\{1.3,1.9\}$. From pa\-nels (a) and (b) one infers that $J$ vanishes for $\phi=\{0,\pm\pi/2,\pi\}$, a result which is fully compatible with the fact that $J$ retains contributions only from the term$\propto\tau_2$ in Eq.~\eqref{eq:Jdef}.

\subsection{Susceptibility analysis}

Another important ingredient for the construction of the BSM is the concept of the generalized susceptibi\-li\-ties which provide the coefficients for the expansion of the response fields for weak variations $(\delta\phi,\delta B)$ about the va\-lues $(\phi,B)$. These variations are supposed to give rise to the spatiotemporally varying part of the external fields, and are employed to implement a spacetime texture in the synthetic space, which is required for the BSM study. Standard Maclaurin expansions for $(\phi,B)$ yield:
\bea
J(\phi+\delta\phi,B+\delta B)&\approx&\chi_{\phi\phi}(\phi,B)\delta\phi+\chi_{\phi B}(\phi,B)\delta B,\no\\
M(\phi+\delta\phi,B+\delta B)&\approx&\chi_{B\phi}(\phi,B)\delta\phi+\chi_{BB}(\phi,B)\delta B,\no
\eea

\noi and the respective susceptibilities are thus defined as:
\bea
\chi_{\phi\phi}(\phi,B)&=&\partial_\phi J(\phi,B),\phd\chi_{BB}(\phi,B)=\partial_B M(\phi,B),\qquad\label{eq:SuscDef1}\\
\chi_{\phi B}(\phi,B)&=&\partial_B J(\phi,B),\phd\chi_{B\phi}(\phi,B)=\partial_\phi M(\phi,B)\,.\qquad
\label{eq:SuscDef2}
\eea

\begin{figure}[t!]
\begin{center}
\includegraphics[width=\columnwidth]{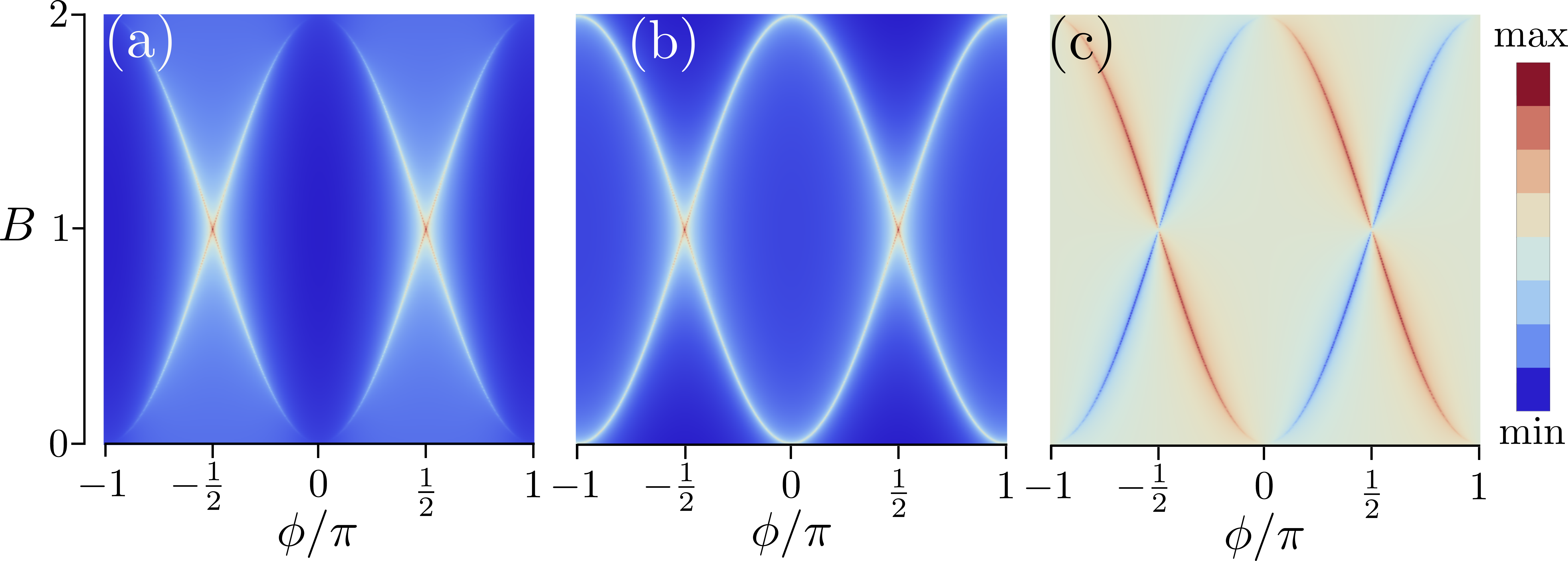}
\end{center}
\caption{Panels (a), (b) and (c) depict heat maps of the nume\-ri\-cal\-ly evaluated susceptibilities $\chi_{\phi\phi,BB,\phi B}(\phi,B)$, respectively, for $\beta=0$. Note that $\chi_{\phi B}=\chi_{B,\phi}$. The {\color{black}colorcoding} is set by the respective bar, where deep red (blue) corresponds to the maximum and minimum of the depicted function.}
\label{fig:Figure3}
\end{figure}

\noi Note that the Onsager-type of reciprocity relation $\chi_{\phi B}=\chi_{B\phi}$ holds for the two mixed susceptibilities as also found in Ref.~\onlinecite{PK}. The results of the numerical evaluations of the various su\-sceptibility elements are presented in Fig.~\ref{fig:Figure3}, and exhibit strong similarities to the ones obtained for the AIII topological insulator model of Ref.~\onlinecite{PK}.

To improve our understanding in regards with the behavior of the susceptibilities, it is useful to consider the evaluation of these quantities using second order perturbation theory in response to the variations $(\delta\phi,\delta B)$. One finds that up to lowest order in these variations, the ope\-ra\-tors providing the response of the system now read:
\bea
\hat{J}(k,\phi+\delta\phi,B+\delta B)&\approx&\hat{J}(k,\phi,B)-\sin\phi\sin k\hat{\mathds{1}}_\tau\delta\phi\no\\
&&-\cos\phi\cos k\tau_2\delta\phi\,,
\label{eq:Jdefvar}\\
\hat{M}(k,\phi+\delta\phi,B+\delta B)&=&\hat{M}(k,\phi,B)\,.
\label{eq:Mdefvar}\eea

\noi The susceptibilities retain two types of contributions from the generalized currents of Eqs.~\eqref{eq:Jdefvar} and~\eqref{eq:Mdefvar}.  The first originates from the linear response to the perturbation: $\delta\hat{H}_{\rm BdG}(k,\phi+\delta\phi,B+\delta B)\approx-\hat{J}(k,\phi,B)\delta\phi-\hat{M}(\phi,B)\delta B$. The presence of the charge-conjugation symmetry $\Xi$ implies that the contributions from the dia\-go\-nal part of the current in Eq.~\eqref{eq:Jdef} drop out completely also when evalua\-ting the respective current-current correlation function, therefore lea\-ving only the term $\propto\tau_2$ to contribute. {\color{black}Hence}, the linear response contribution to the susceptibilities of the system is associated with the quantum metric~\cite{Provost} of the occupied band and, in turn, with the Berry vorticity density which corresponds to the integrand of Eq.~\eqref{eq:Vorticity}. {\color{black}A similar result was found} pre\-viou\-sly in Ref.~\onlinecite{PK}. Therefore, this so-called geometric part of the response, provides direct information regar\-ding the di\-stan\-ce from a Berry sin\-gu\-la\-ri\-ty and the related to\-po\-lo\-gi\-cal phase transition~\cite{CPSun,Zanardi,HQLin,Ma1,Matsuura,Ma2}. The second contribution to the susceptibilities in unrelated to the quantum geo\-metry and, as discussed in Ref.~\onlinecite{PK}, it is practically constant along the Berry singularity branches and subdo\-mi\-nant to the corresponding geometric contribution. This is because, this nongeometric part stems from the expectation value of the se\-cond derivative of the currents $\hat{J}(k,\phi+\delta\phi,B+\delta B)$ and $\hat{M}(k,\phi+\delta\phi,B+\delta B)$. Here, it is only the former term that yields such a contribution.

\begin{figure}[t!]
\begin{center}
\includegraphics[width=\columnwidth]{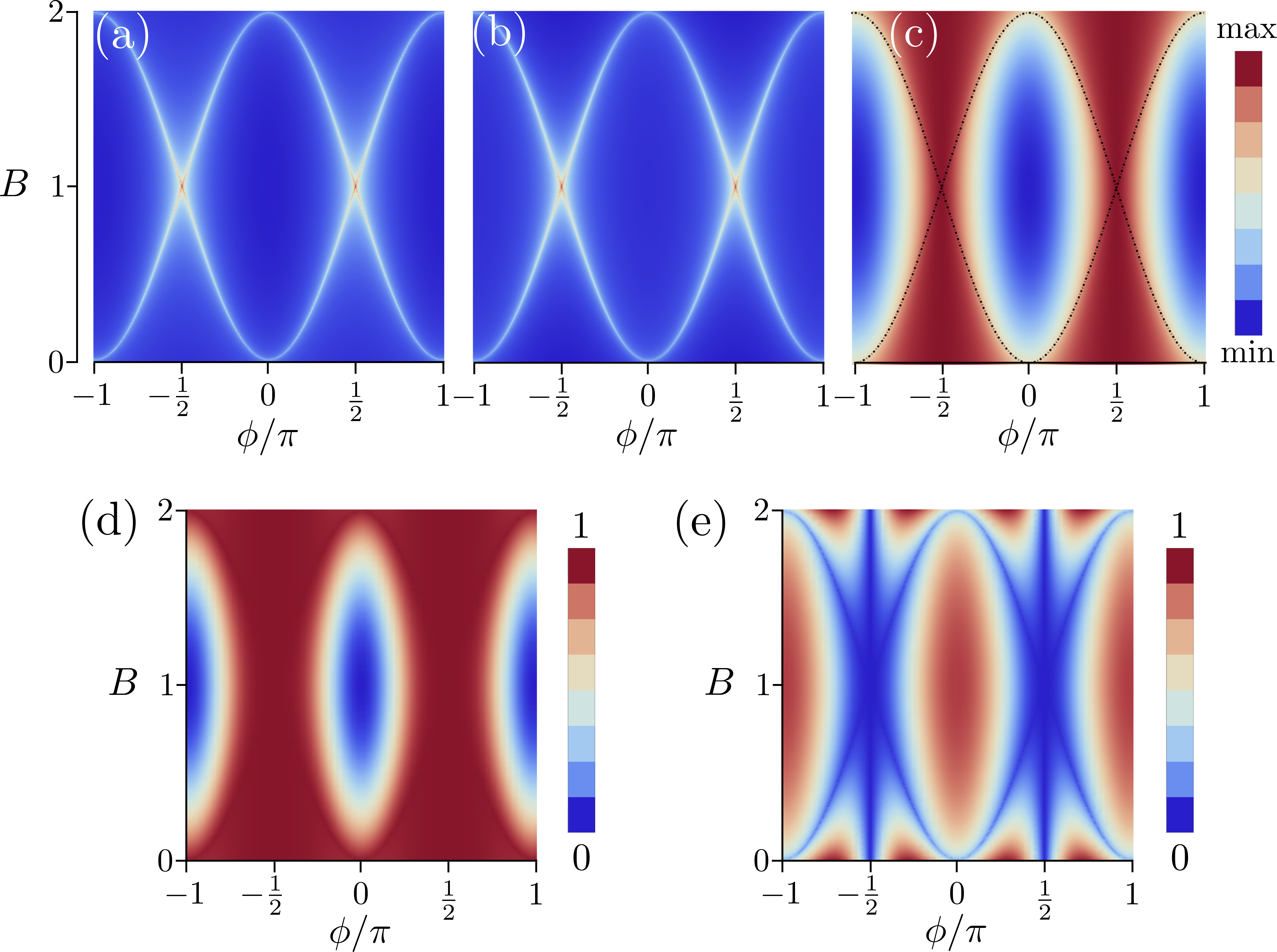}
\end{center}
\caption{Numerical evaluations of (a) the trace ${\rm Tr}(\check{\chi})=\chi_{\phi\phi}+\chi_{BB}$ of the total susceptibility tensor, (b) its geo\-me\-tric part ${\rm Tr}(\check{\chi}_1)$, (c) its nongeometric part ${\rm Tr}(\check{\chi}_2)$, and (d/e) the normalized geometric/nongeometric parts $|{\rm Tr}(\check{\chi}_{1,2})|/\sqrt{[{\rm Tr}(\check{\chi}_1)]^2+[{\rm Tr}(\check{\chi}_2)]^2}$. From panels (a)-(b) one confirms that the total trace of $\check{\chi}$ and its geometric part are almost identical and become sizeable in the vicinity of the Berry singularities. In contrast, the nongeometric part is significant and dominant away from the Berry singularity branches.}
\label{fig:Figure4}
\end{figure}

The above conclusions are confirmed by plotting the separate contributions to the susceptibility elements. Of particular interest is the trace ${\rm Tr}(\check{\chi})=\chi_{\phi\phi}+\chi_{BB}$ of the susceptibility tensor $\check{\chi}$, since this enters the evaluation of the BSM. Figure~\ref{fig:Figure4} depicts the evaluation of the above trace for the total susceptibility, as well as its geo\-metric ${\rm Tr}(\check{\chi}_1)$ and nongeometric ${\rm Tr}(\check{\chi}_2)$ parts. The total and geometric contributions behave similarly, while the nongeometric part exhibits a distinct dependence on the parameters $(\phi,B)$. Panel (c) corroborates that the nongeometric part is practically constant in the vicinity of the Berry singularity branches. Panels (d) and (e) depict the normalized geo\-me\-tric and nongeometric contributions, and clearly demonstrate that the geometric contribution is dominant is a broad region centered at the Berry singularity branches.

Concluding this analysis, I wish to mention that a number of recent works, cf Refs.~\onlinecite{Torma,Rossi,BAB}, have also discussed the role of the quantum metric in the calculation of the superfuid density, which is related to the evaluation of $J$ here. However, the reasons for the emergence of quantum geometry in those works and the present one are radically different. The former focus on multiband SCs and the supercurrent carried by flat bands. As they have shown, the supercurrent carried by a flat band is not only nonzero, as one may naively expect, but it is even bounded from below~\cite{Torma,Rossi,BAB}. This is due to contributions ori\-gi\-na\-ting from interband transitions, thus naturally involving the quantum metric. Here instead, I consider a single-band $p$-wave SC, whose nontrivial dispersion is solely responsible for the emergence of the quantum metric.

\subsection{Berry singularity marker}

The results obtained for the response fields and the susceptibilities are very similar to the ones encountered in the topological insulator model of Ref.~\onlinecite{PK}. Therefore, one expects to obtain an equally similar behaviour for the BSM, which is here defined compactly as:
\bea
&&{\cal Q}(\phi,B,{\cal B})=\int_0^{2\pi}\frac{d\omega}{\pi}\Big[\cos\omega{\cal A}_B(\phi,B,{\cal B},\omega)\quad\quad\no\\&&\qquad\qquad\qquad\qquad\quad\quad-\sin\omega{\cal A}_\phi(\phi,B,{\cal B},\omega)\Big]\,,
\label{eq:BSM}
\eea

\noi with the help of the two quantities:
\bea
{\cal A}_\phi(\phi,B,{\cal B},\omega)&=&-\frac{{\rm sgn}\left[\chi_{\phi B}(\phi,B)\right]}{{\rm Tr}\big(\check{\chi}_{\rm BS}\big){\cal B}}M(\phi+\delta\phi,B+\delta B)\,,\no\\\\
{\cal A}_B(\phi,B,{\cal B},\omega)&=&+\frac{{\rm sgn}\left[\chi_{\phi B}(\phi,B)\right]}{{\rm Tr}\big(\check{\chi}_{\rm BS}\big){\cal B}}J(\phi+\delta\phi,B+\delta B)\,.\no\\\eea

\noi The above play a role analogous to the Berry vector potential of the occupied/empty bands, and are obtained after considering a spacetime helix for the variations: $\delta\phi={\cal B}\cos\omega$ and $\delta B={\cal B}\sin\omega$, which is parametrized by the phase $\omega\in[0,2\pi)$. The quantity ${\rm Tr}\big(\check{\chi}_{\rm BS}\big)$ denotes the value of ${\rm Tr}\big(\check{\chi}\big)$ eva\-lua\-ted at the location of the Berry singularity which is the closest to the synthetic space location $(\phi,B)$ of interest. {\color{black} The emergence of ${\rm Tr}\big(\check{\chi}_{\rm BS}\big)$ in the definition of the BSM becomes transparent when con\-si\-de\-ring that ${\cal B}$ is sufficiently small, so to justify the Maclaurin expansion of the response fields $M(\phi+\delta\phi,B+\delta B)$ and $J(\phi+\delta\phi,B+\delta B)$ which was discussed in Sec.~\ref{sec:SectionIII}. Under this condition, and for the above assumed spacetime helix profile for $\delta\phi$ and $\delta B$, the expression for the BSM simplifies as:
\bea
&&{\cal Q}_{{\cal B}\rightarrow0}(\phi,B)=\frac{{\rm sgn}\left[\chi_{\phi B}(\phi,B)\right]}{{\rm Tr}\big(\check{\chi}_{\rm BS}\big)}\int_0^{2\pi}\frac{d\omega}{\pi}\Big[\cos^2\omega\chi_{\phi\phi}(\phi,B)\no\\&&\ph\qquad\qquad+\sin^2\omega\chi_{BB}(\phi,B)+\sin(2\omega)\chi_{\phi B}(\phi,B)\Big]\,.
\label{eq:BSMlimit}
\eea

\noi The integrals appearing above are elementary and directly yield that, in this limit, the BSM is given by:
\begin{align}
{\cal Q}_{{\cal B}\rightarrow0}(\phi,B)={\rm sgn}\left[\chi_{\phi B}(\phi,B)\right]{\rm Tr}\big[\check{\chi}(\phi,B)\big]/{\rm Tr}\big(\check{\chi}_{\rm BS}\big)\,. 
\end{align}

\noi The above sheds light to the construction of the BSM as well as the reason for introducing ${\rm Tr}\big(\check{\chi}_{\rm BS}\big)$, which functions as a normalization constant, in order to ensure that the BSM becomes equal to $\pm1$ when the values of $(\phi,B)$ are tuned at the BS. Moreover, this result further reveals that the contribution of the mixed susceptibility $\chi_{\phi B}(\phi,B)$ drops out after the integration. Hence, the BSM receives only contributions from the diagonal response for ${\cal B}\rightarrow0$. Note, however, that this holds as long as $(\delta\phi,\delta B)$ form a spacetime helix as above, while any ``disorder'' from this ideal profile renders the contribution of the mixed response non-negligible.
}

Given the above, it is straightforward to numerically evaluate the BSM for the model of Eq.~\eqref{eq:BdGHamiltonian} under the assumption of $\beta=0$. For the numerical implementation, a discrete helix is considered consisting of $L$ points. In this case, the expression for ${\cal Q}$ has to be modified using the replacement $\int_0^{2\pi}d\omega/2\pi\mapsto\nicefrac{1}{L}\sum_n$, where $n=1,2,\ldots,L$ labels the points of the helix. 

\begin{figure}[t!]
\begin{center}
\includegraphics[width=0.95\columnwidth]{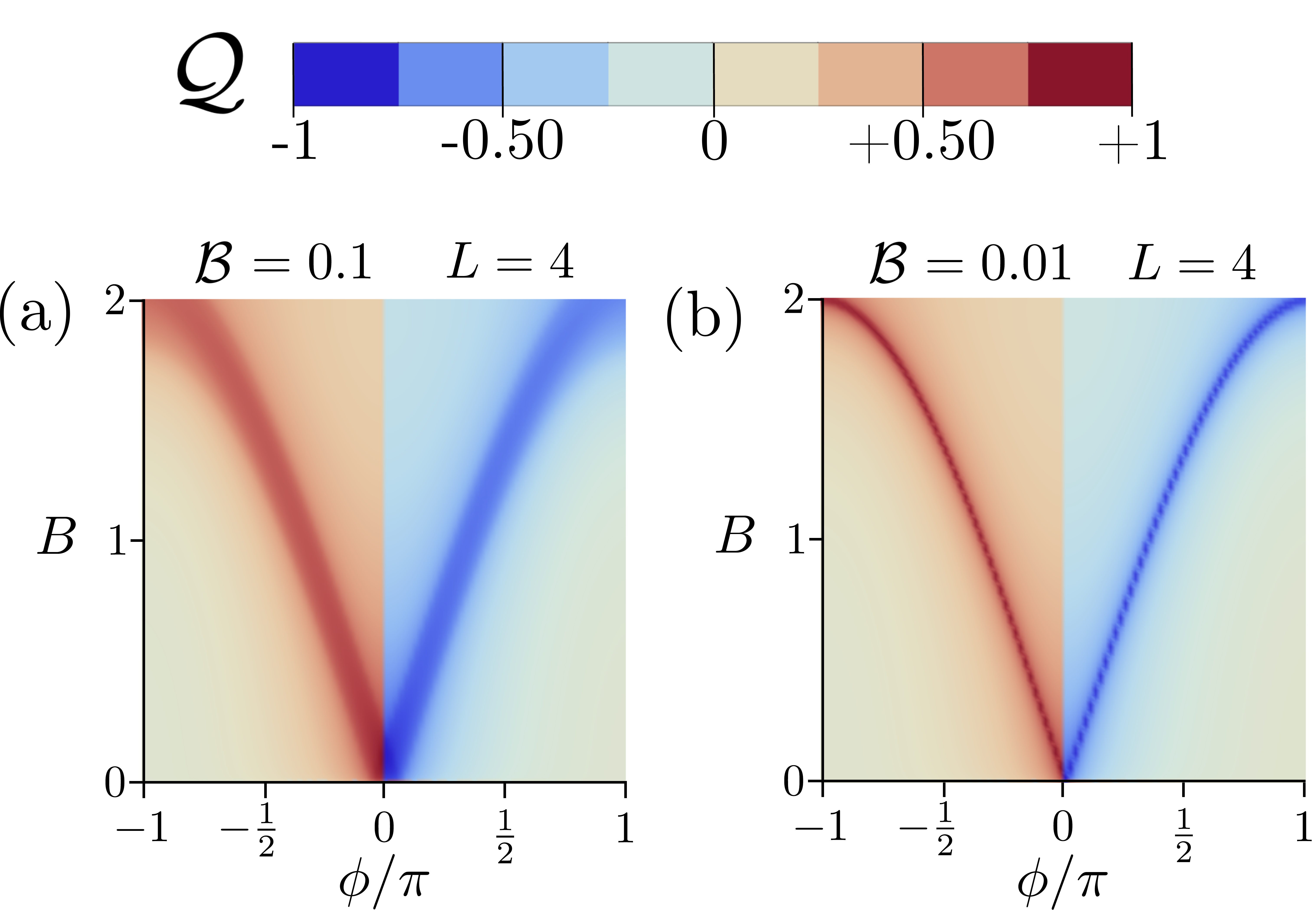}
\end{center}
\caption{Numerically evaluated BSM. The panels depict a heat map of the BSM value ${\cal Q}$, according to the colorco\-ding presenting on the top of the fi\-gu\-re. Panels (a) and (b) are calculated for dif\-fe\-rent resolution given by the values ${\cal B}=\{0.1,\,0.01\}$ and the same lengths $L=4$ of the spacetime helix $(\delta\phi,\delta B)={\cal B}(\cos\omega,\sin\omega)$. In the calculations, ${\rm Tr}\big(\check{\chi}_{\rm BS}\big)$ was replaced by ${\rm Tr}\big\{\check{\chi}\big(\phi=\cos^{-1}\big(B-1\big),B\big)\big\}$, i.e., the value obtained for the Berry singularity closest to $(\phi,B)$.}
\label{fig:Figure5}
\end{figure}

Figure~\ref{fig:Figure5} shows related results for fixed $L=4$, and the two values $\{0.1,0.01\}$ for the parameter ${\cal B}$, which quantifies the resolution of the experimental probe. From the results of Fig.~\ref{fig:Figure5} one confirms that the decrease of ${\cal B}$, which implies the enhancement of the experimental re\-so\-lu\-tion, leads to higher BSM re\-so\-lution and can better discern the Berry singularities. As shown in Ref.~\onlinecite{PK}, increasing the length or the number of sampling cycles is meaningful when disorder is present, and it provides a path to render the evaluation of the BSM immune to disorder. Due to the strong similarities {\color{black}obtained for the BSM results} of the present model and the one examined in  Ref.~\onlinecite{PK}, I do not carry out here investigations in the presence of disorder.

\section{Extended p-wave SC Model}\label{sec:SectionIV}

In this section, I proceed with the investigation of a number of extensions considered for the standard model, which become relevant in the presence of chiral symmetry. Specifically, by switching on $\beta$, additional pairs of Berry singularities appear, while gapless phases become accessible. The latter obstruct the evaluation of the BSM defined using the flux, and call for alternative BSMs.

\subsection{Topological phases and Berry singularities}

\begin{figure}[t!]
\begin{center}
\includegraphics[width=\columnwidth]{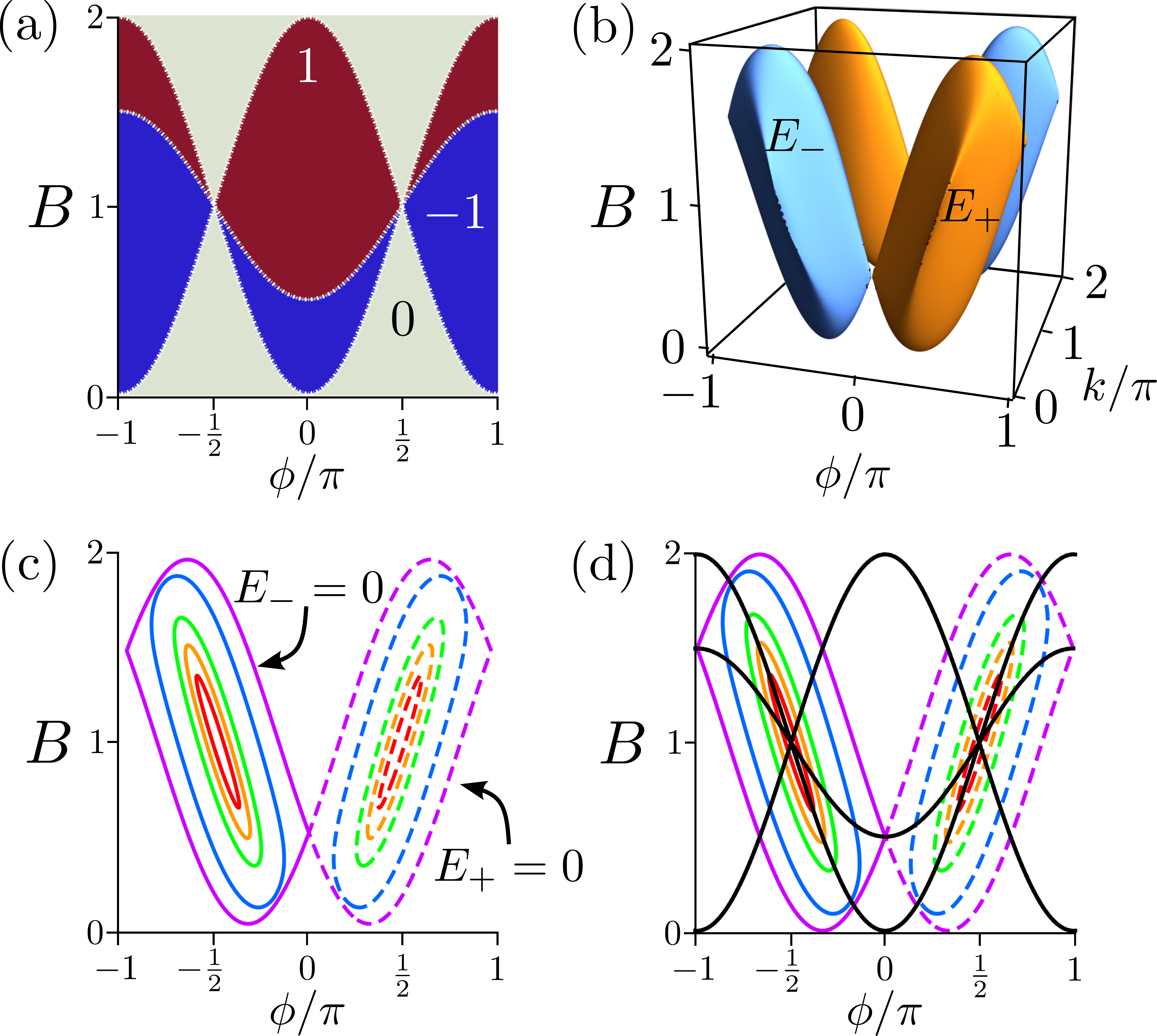}
\end{center}
\caption{(a) Winding number value diagram. One finds a tri\-vial ($0$) and two nontrivial ($\pm1$) values. The winding number is ill-defined for $B=1\pm\cos\phi$ and $B=1-\cos\phi/\beta$, where band touchings occur at $k=\{0,\pi,\pm k_*\}$, where $\beta\cos k_*=1$. Note that this diagram is capable of predicting the to\-po\-lo\-gi\-cal properties, only as long as the bulk energy spectrum is fully gapped. Here, gapless phases arise for fluxes $\phi\neq\{0,\pi\}$. Pa\-nels (b), (c) and (d) exemplify the emergence of band cros\-sings in $k$ space, Bogoliubov-Fermi surfaces in the mixed synthetic 3D space $(\phi,k,B)$, and Bogoliubov-Fermi loops in the pure synthetic 2D space $(\phi,B)$. The last panel illustrates how the gapless phases arise from touchings of the $k=\{0,\pi,\pm k_*\}$ points. In the present analysis $\beta=2$.} 
\label{fig:Figure6}
\end{figure}

In this paragraph, I carry out the topological classification for $\beta\neq0$. When $\phi=\{0,\pi\}$ the Hamiltonian lies in class BDI for arbitrary nonzero va\-lues chosen for $\beta$ and $B$. Fi\-gu\-re~\ref{fig:Figure6}(a) shows the possible topological phases of the model under consideration for $\beta=2$. These are inferred by evaluating the winding number defined in Eq.~\eqref{eq:WindingNumber} as a function of $\phi\in[-\pi,\pi)$ and $B\in[0,2]$. Besides the case of a trivial winding, one finds phases with the nonzero winding numbers $w=\pm1$. The occu\-ring winding number transitions appear now along three paths in $(\phi,B)$ space, and are associated with the gap closings at $k=0$, $k=\pi$, as well as two additional points given by $\beta\cos k_*=1$. The gap closings at $k=\pm k_*$ give rise to the Berry singularity branch $B=1-\cos\phi/\beta$. The chiral symmetry of the BdG Hamiltonian, which is effected by $\Pi=\tau_3$, implies that the Berry sin\-gu\-la\-ri\-ties $k=\pm k_*$ appear for identical $(\phi,B)$, and possess the same to\-po\-lo\-gi\-cal charge $Q$. Employing Eq.~\eqref{eq:Vorticity} for $\phi=0$, yields $Q=\{-1,-1,1\}$ for $k=\{0,\pi,\pm k_*\}$, respectively. Remarkably, the long-ranged pairing results in the switching of the sign of the Bery singu\-la\-ri\-ty at $k=\phi=B=0$, compared to the one found in Sec.~\ref{Sec:TopoPhasesStandard}.

Class D Hamiltonians are accessible for $\phi\neq\{0,\pi\}$, as shown in the previous section. As mentioned earlier, the presence of flux does not invalidate the winding number which can be still employed to classify the topological phases of the system. However, in the present case $\beta\neq0$ which implies that gapless phases set in. In the regions of the synthetic space $(\phi,B)$ where gapless phases develop, the BSM defined earlier does not any longer constitute a good tool to detect Berry singularities, due to the pre\-sen\-ce of the crossing points which obscure the outcome.

In Figs.~\ref{fig:Figure6}(b),~(c), and~(d) I discuss the emergence of gapless phases which arise for $\phi\neq\{0,\pi\}$. In panel (b), I present a countour plot of the Fermi surface obtained in the 3D $(\phi,k,B)$ mixed synthetic space. The emergence of Bogoliubov-Fermi surfaces~\cite{ChiuRMP} restricts the validity of the winding number to a narrow parameter value window, and at the same time reduces the capabilities of the flux knob to probe the topological pro\-per\-ties in the insulating regime. Panels (c) and (d) provide Fermi surface cuts of (b) in $(\phi,B)$ space. The contours shown from inside to outside correspond to wavenumbers $k=\{\pi/12,\pi/8,\pi/6,\pi/4,\pi/3\}$. One observes that a flux can be employed as a noninvasive tool to infer the topological properties of the gapped phase only for a neighborhood of values centered at $(\phi,B)=(0,\nicefrac{3}{2})$ and $(\phi,B)=(\pi,\nicefrac{1}{2})$. In these regions, the insulating phase persists and the winding number analysis remains valid.

\subsection{Response to the external fields}

In this paragraph I investigate the modifications that the presence of the added longer-ranged pariring term incurs on the response fields $J$ and $M$. The numerical evaluation of the various quantities is considered at zero temperature which I obtain here by considering a finite temperature formalism. The inclusion of a nonzero temperature $T$ (in energy units), is here used to smoothen out various numerical instabilities appearing when using instead the zero-temperature formalism. The evaluation is carried out by defining the fermionic free energy part:
\begin{align}
F(\phi,B)=-T\sum_{s=\pm}\int_{\rm BZ}\frac{dk}{2\pi}\ln\left[1+e^{-E_s(\phi,B)/T}\right]
\end{align}

\noi and subsequently defining $J=-\partial_\phi F $ and $M=-\partial_BF $.

\begin{figure}[t!]
\begin{center}
\includegraphics[width=\columnwidth]{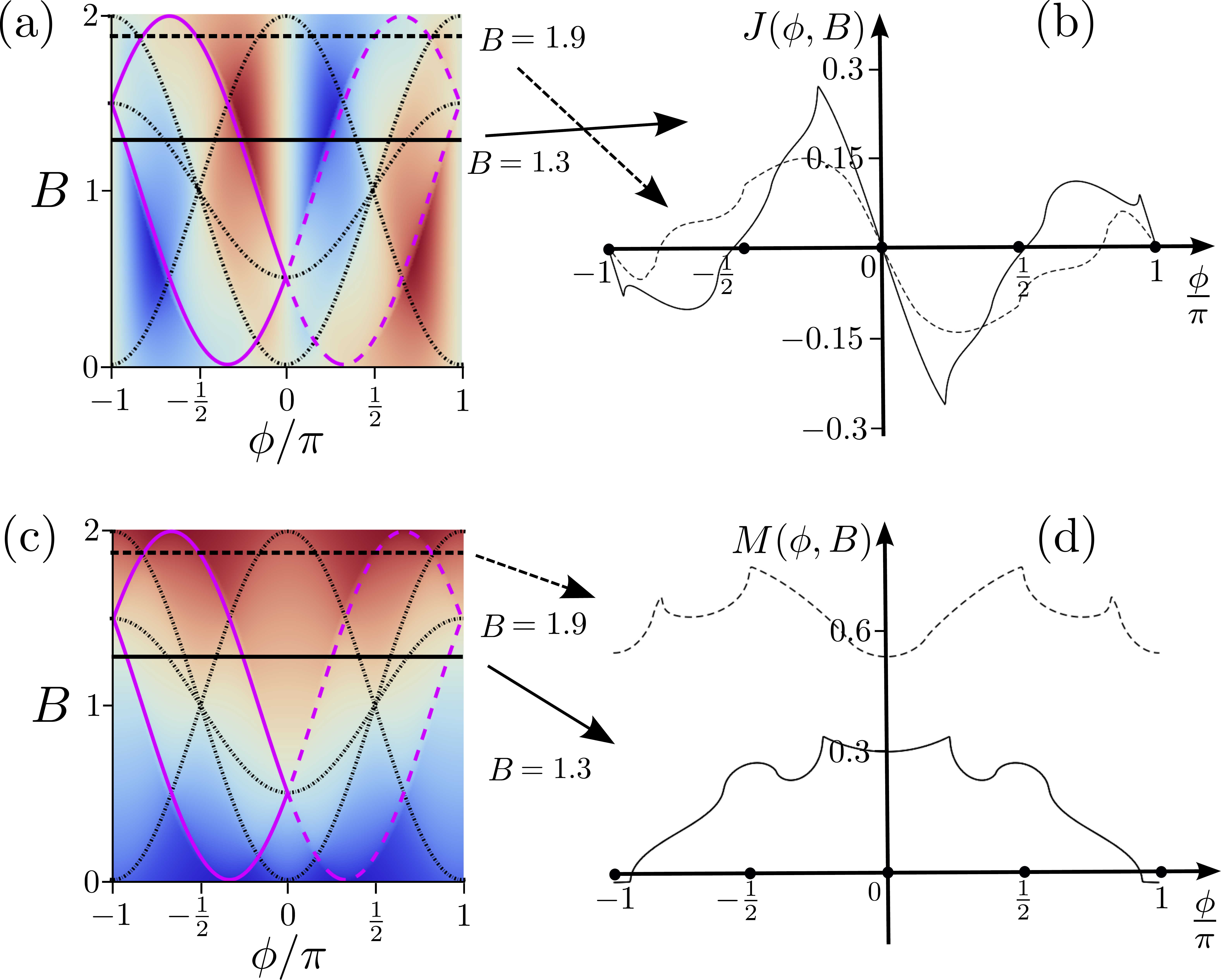}
\end{center}
\caption{Numerical evaluation of $(J,M)$ as functions of $(\phi,B)$. Panels (a) and (c) depict heat maps for $(J,M)$, with a color\-co\-ding which varies continuously between deep red to deep blue (\includegraphics[scale=0.06]{BSM_TSC_0}), with these two values corresponding to the maximum and minimum of the depicted function. Pa\-nels (b) and (d) present cuts of (a) and (c) for $B=1.9$ and $B=1.3$. In the heat maps I have superimposed the three Berry singularity paths (black dot-dashed) $B=1\mp\cos\phi$ and $B=1-\cos\phi/\beta$, as well as a countour (magenta) which determines the parameter values of $(\phi,B)$ for which a Bogoliubov-Fermi loop is formed at the wavenumber $k=\pi/3$. As shown in Fig.~\ref{fig:Figure6}(c), the Bogoliubov-Fermi loop obtained for $k=\pi/3$ in the 3D space $(\phi,k,B)$ yields the Fermi countour with the largest area in the 2D $(\phi,B)$ plane. Hence, the area enclosed in the magenta contour captures the part of the heat map for which, a gapless phase is stabilized. Both conjugate fields feature discontinuous slopes for the values of $\phi$ where band crossings appear. In the above, I set $\beta=2$ and $T=2\cdot10^{-3}$.}
\label{fig:Figure7}
\end{figure}

A re\-la\-ted nu\-me\-ri\-cal analysis resulting from the above procedure is presented in Fig.~\ref{fig:Figure7}. One observes that the main effect of the additional pairing term is to introduce the gapless phases presented in Fig.~\ref{fig:Figure6}(d). When these set in, the response fields become substantially modified, cf Fig.~\ref{fig:Figure7}(b)~and~(d). This is because the (free)energy now obtains/loses contributions from the upper/lower band. Two important conclusions arise from Fig.~\ref{fig:Figure7}. First, the gapless phases can be detected through discontinuities in the slope of the response which appear when these set in and, second, the branches of the pairs of Berry sin\-gu\-la\-ri\-ties appearing at $k=\pm k_*$, are completely buried inside the gapless phases. Hence, the Berry singularity pairs cannot be accessed by the BSM defined in Eq.~\eqref{eq:BSM} in terms of $(J,M)$. On the other hand, the abovementioned BSM is still capable of diagnosing the Berry singularities stemming from gap closings at $k=\{0,\pi\}$, but now in a significantly reduced window of the synthetic space.  

\subsection{BSM defined in strain - magnetic field space}

In order to tackle the inaccessibility of the pair of Berry singularity pairs by the so far examined BSM, as well as enable the study of their interplay with the Berry singularities emerging from $k=\{0,\pi\}$, one can consider alternative BSMs. To identify a more suitable BSM for this purpose, one needs to replace the flux knob by a field which provides control over the phase transition me\-dia\-ted by the pair of Berry singularities. From the structure of the Hamiltonian one infers that $\beta$ is a natural parameter candidate. Indeed, the positions of these Berry singularities are tunable by modifying $\beta$, since $\cos k_*=1/\beta$. In the remainder of this section, I construct a BSM in terms of $(\beta,B)$ and set $\phi=0$. The parameter $\beta$ is dimensionless in all unit systems, and its size can be controlled by effecting strain along the main axis of the sample. The conjugate field of $\beta$ corresponds to an elastic modulus, i.e., a stress field which I here denote $\sigma$.

\begin{figure*}[t!]
\begin{center}
\includegraphics[width=\textwidth]{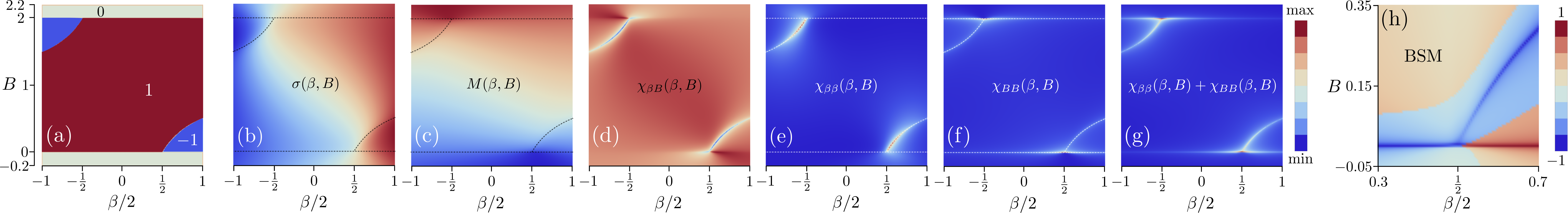}
\end{center}
\caption{(a) Topological phase diagram of the Hamiltonian in Eq.~\eqref{eq:BdGHamiltonian} for $\phi=0$. (b) and (c) Numerical evaluation of $(\sigma,M)$ as functions of $(\beta/2,B)$. Panels (d)-(f) depict the susceptibility elements $\chi_{\beta B,\beta\beta,BB}(\beta,B)$, while (g) shows the trace $\chi_{\beta\beta}(\beta,B)+\chi_{BB}(\beta,B)$. Notably, the generalized susceptibilities retain only geometric contributions. (h) shows the BSM value ${\cal Q}(\beta,B)$. At $(\beta/2,B)=(1/2,0)$ two different Berry singularity branches coalesce. To evaluate the BSM, I set ${\rm Tr}\big(\check{\chi}_{\rm BS}\big)={\rm Tr}\big\{\check{\chi}\big(\beta,B=0\big)\big\}$.}
\label{fig:Figure8}
\end{figure*}

Figure~\ref{fig:Figure8} presents the to\-po\-lo\-gi\-cal phase diagram and the responses of the TSC in terms of $(\sigma,M)$. Varying $\beta$ for a fixed $B$ and $\phi=0$, allows observing the transition $w=1\leftrightarrow w=-1$ of Fig.~\ref{fig:Figure6}(a). Since the system resides in the BDI class, only the $E_-$ band determines the response and, thus no gapless phases arise. The su\-sceptibilities peak as usual in the vicinity of the Berry singularity branches. From Fig.~\ref{fig:Figure8}(a), one observes that there exist two points in the $(\beta,B)$ space where two Berry sin\-gu\-la\-ri\-ty branches meet. Figure~\ref{fig:Figure8}(h) depicts the result of BSM defined in $(\beta,B)$ space. One observes that at $(\beta/2,B)=(1/2,0)$ the pair of Berry singularities car\-rying here a total charge of $-2$ units coalesce with the Berry singularity branch stemming from the $k=0$, which extends along $(\beta/2>1,B=0)$ and sees a charge $Q=1$. Their merging gives rise to the Berry singularities for $(\beta/2<1,B=0)$ with a charge $-1$. It is important to note that in the present case it is not possible to identify the topological charges of the various Berry singularities in a precise manner, since the expansion of the Hamiltonian about $k=0$ is not linear in neither $k$ nor $B$.

\section{Conclusions and outlook}\label{sec:SectionV}

To summarize, I described how to {\color{black}extend the BSM method to the case of $p$-wave SC models} which belong to symmetry class BDI. The results obtained for the standard Kitaev chain model are similar to the ones found in Ref.~\onlinecite{PK} for a AIII topological insulator model. However, here the external field employed to mimic the wavenumber $k$ is a flux knob which induces a supercurrent in the SC. The flux field can also detect Berry singularities for SCs with longer-ranged pairing, but it has restricted capabilities when it comes to the detection of pairs of Berry singularities. To tackle this shortcoming I introduced an alternative BSM defined using a strain field instead.

As it became apparent in this work, a number of aspects of the BSM approach need to be further con\-si\-de\-red. Firsty, this method needs to be generalized and syste\-ma\-ti\-zed. Second, additional effort is needed to address how to efficiently tell Berry singularities of different charges apart based on the quantum geometry. Concerning the experimental implementation of this method, the BSM approach appears feasible to be realized in semiconductor nanowire hybrids~\cite{LutchynPRL,OregPRL} by consi\-de\-ring the proximity to conventional SC segments which are parts of larger SC rings which are pierced by flux in a controllable fashion.

\section*{Acknowledgements}

I thank Kun Bu for helpful discussions on related topics which inspired part of this work. Moreover, I acknowledge funding from the National Na\-tu\-ral Scien\-ce Foundation of China (Grant No.~12074392 and No.~12050410262).

\end{document}